%% file: main.tex
\documentclass[runningheads]{llncs}
\usepackage{multirow}
\usepackage{enumitem}
\setlist{nosep}

\usepackage[T1]{fontenc}

\usepackage{hyperref}
\usepackage{color}

\usepackage{graphicx}
\usepackage{amsmath}
\usepackage{booktabs}
\usepackage{multirow}
\usepackage[table]{xcolor} 
\usepackage{etoolbox}

\definecolor{BiasBlue}{HTML}{D6EAF8}
\definecolor{BiasOrange}{HTML}{FDEBD0}

\begin{document}

\hypersetup{
  pdftitle={Training-Induced Bias Toward LLM-Generated Content in Dense Retrieval},
  pdfauthor={Your Real Name et al.}, 
  pdfcreator={LaTeX},
  pdfproducer={Springer}
}
\title{Training-Induced Bias Toward LLM-Generated Content in Dense Retrieval}

\author{William Xion \and Wolfgang Nejdl}
\authorrunning{W. Xion and W. Nejdl}
\institute{L3S Research Center, Hannover, Germany\\
\email{william.xion@l3s.de, nejdl@L3S.de}}
\maketitle              

\begin{abstract}
Dense retrieval is a promising approach for acquiring relevant context or world knowledge in open-domain natural language processing tasks and is now widely used in information retrieval applications. However, recent reports claim a broad preference for text generated by large language models (LLMs). This bias is called "source bias", and it has been hypothesized that lower perplexity contributes to this effect. In this study, we revisit this claim by conducting a controlled evaluation to trace the emergence of such preferences across training stages and data sources. Using parallel human- and LLM-generated counterparts of the SciFact and Natural Questions (NQ320K) datasets, we compare unsupervised checkpoints with models fine-tuned using in-domain human text, in-domain LLM-generated text, and MS MARCO. Our results show the following: 1) Unsupervised retrievers do not exhibit a uniform pro-LLM preference. The direction and magnitude depend on the dataset. 2) Across the settings tested, supervised fine-tuning on MS MARCO consistently shifts the rankings toward LLM-generated text. 3) In-domain fine-tuning produces dataset-specific and inconsistent shifts in preference. 4) Fine-tuning on LLM-generated corpora induces a pronounced pro-LLM bias. Finally, a retriever-centric perplexity probe involving the reattachment of a language modeling head to the fine-tuned dense retriever encoder indicates agreement with relevance near chance, thereby weakening the explanatory power of perplexity. Our study demonstrates that source bias is a training-induced phenomenon rather than an inherent property of dense retrievers.\footnote{Code is available at \url{https://github.com/williamx854/finetuning-source-bias}.}

\end{abstract}
%
\input{introduction}
\input{related_work}
\input{experiments}

\input{results_and_analysis}
\input{conclusion}

\begin{credits}
\subsubsection{\ackname} 
This work was supported by Soofi.

\subsubsection{\discintname}
The authors have no competing interests to declare that are relevant to the content of this article.
\end{credits}

\bibliographystyle{splncs04}
\bibliography{references}

\end{document}

%% file: introduction.tex
\section{Introduction}
\label{sec:introduction}

Dense retrievers are a promising approach in modern information retrieval, designed to capture semantic similarity beyond exact lexical overlap. In recent years, substantial progress has been made in training methodologies, ranging from unsupervised contrastive pre-training to balanced, topic-aware sampling \cite{DBLP:conf/sigir/BalanceAwareLYLH21,DBLP:journals/tmlr/IzacardCHRBJG22}. Consequently, dense models frequently outperform traditional lexical methods, 
such as BM25, in semantic search and question answering benchmarks \cite{DBLP:conf/eacl/MTEBMuennighoffTMR23}.

Alongside these advances, concerns have emerged about bias in dense retrievers. Recent work by Dai et al. \cite{DBLP:conf/kdd/DaiNeuralZPLHLZW024} identified a bias called \textit{source bias} wherein neural retrievers tend to prefer LLM-generated text over semantically equivalent human-authored text. Building on this finding, Wang et al. \cite{DBLP:conf/iclr/WangPerplexityDZP0WD0W25} hypothesized that this bias stems from models inheriting a preference for text with low perplexity, since LLM text typically has lower perplexity \cite{DBLP:conf/iclr/BaoZTY024,DBLP:conf/icml/Mitchell0KMF23}. 

However, existing studies \cite{DBLP:conf/acl/DaiCocktailLZPRWDXW24,DBLP:conf/kdd/DaiNeuralZPLHLZW024,DBLP:conf/iclr/WangPerplexityDZP0WD0W25} examine dense retrievers exclusively in their final, trained state. This leaves unexamined the questions of whether the observed bias represents an inherent characteristic of dense retrievers or a result of the fine-tuning process. This distinction is important because fine-tuning on corpora such as MS MARCO\cite{DBLP:conf/nips/msmarcoNguyenRSGTMD16} has become the de facto standard for achieving state-of-the-art results. However, the mechanism by which such training influences bias is not well understood. Furthermore, the perplexity-based explanation relies on static measurements from general language models (e.g., BERT-base), which are independent of the specific dense retriever being evaluated. Thus, it provides an insufficient explanation for retriever-specific biases that vary across different training configurations and datasets.

In this paper, we revisit the question of whether dense retrievers are inherently biased. We systematically evaluate dense retrievers at different training stages, examining models in their unsupervised state, after general-purpose fine-tuning on MS MARCO, and following in-domain fine-tuning on both human-authored and LLM-generated corpora. Using established benchmarks, we track the evolution of bias throughout the training stages. Additionally, we test the perplexity hypothesis by implementing a retriever-centric perplexity measurement approach that directly reflects the internal representations learned by fine-tuned models.
Our main contributions are as follows:
\begin{itemize}[nosep]
    \item[\textbullet] We provide empirical evidence that unsupervised dense retrievers exhibit inconsistent, dataset-dependent biases. This demonstrates that pro-LLM preference is not an inherent characteristic of these models.
    \item[\textbullet] We identify supervised fine-tuning as the primary driver of source bias and demonstrate that its effect critically depends  on the training corpus. Fine-tuning on general-purpose datasets, such as MS MARCO, consistently induces pro-LLM bias. In contrast, in-domain fine-tuning produces variable preferences depending on the characteristics of the dataset.
    \item[\textbullet] We present a negative result that challenges the prevailing perplexity-based explanation for this phenomenon. Our analysis shows that perplexity does not consistently correlate with retriever preferences, even when measured using retriever-centric approaches that reflect model-specific internal representations.
\end{itemize}

%% file: related_work.tex
\section{Related Work}
\label{sec:related_work}

\subsection{AI-Generated Content and Retrieval Preference}
The growing dominance of AI-generated content (AIGC) has prompted urgent questions about how retrieval systems handle machine-written text. The investigation into this area for dense retrievers started with Dai et al. \cite{DBLP:conf/kdd/DaiNeuralZPLHLZW024}, who coined the term \textit{source bias}. They demonstrated that dense retrievers favor LLM-generated passages over semantically equivalent human text. Building on this, Wang et al. \cite{DBLP:conf/iclr/WangPerplexityDZP0WD0W25} sought to explain this phenomenon, hypothesizing that models implicitly use perplexity as a relevance signal. This preference is not limited to text; subsequent work has shown that the bias extends to other modalities. Xu et al. \cite{DBLP:conf/sigir/Xu_imageHPDXSC24} found text-image retrieval models prefers AI-generated images, while Gao et al. \cite{DBLP:journals/corr/Gao_abs-2502-07327} found a similar preference for AI-generated videos. These studies collectively illustrate a concerning pattern in which retrieval systems may prefer AIGC. Our work examines when this bias emerges and how it is influenced by fine-tuning.

\subsection{Broader Biases in Retrieval Systems}
Beyond a preference for AI-generated content, research has shown that dense retrievers are susceptible to a range of other biases, often by learning unintended "shortcuts." For example, Fayyaz et al. \cite{DBLP:conf/acl/FayyazMS025} showed that dense retrievers often prioritize passages based on surface-level cues like document length or term position rather than pure semantic relevance, while Cao et al. \cite{DBLP:conf/wsdm/Cao25} found they favor specific writing styles such as formal text. Other work has uncovered more latent issues, showing that retriever embeddings can encode demographic attributes like gender \cite{DBLP:journals/corr/Goldfarbabs-2402-15925}, or that their effectiveness can vary with the linguistic complexity of a query \cite{DBLP:conf/naacl/ChengA25}. While these studies confirm that dense retrievers readily learn surface-level 
shortcuts, recent work \cite{DBLP:conf/iclr/WangPerplexityDZP0WD0W25} suggests that the preference for 
AI-generated content is another instance of such shortcut learning, arising from 
dense retrievers' preference towards lower perplexity passages. Our work further 
investigates this claim by tracking the evolution of this preference across 
distinct training stages.

%% file: experiments.tex
\section{Experimental Setup}
\label{sec:experimental_setup}

\subsection{Research Questions}

To examine whether dense retrievers exhibit consistent preferences for LLM-generated content, we evaluate models across different stages of training and when fine-tuned on different datasets. While prior studies identified a prevalent pro-LLM bias, they largely focused on dense retrievers in their final, fully fine-tuned state. We examine whether this preference emerges consistently across training stages or varies depending on the fine-tuning process and training data composition. To test this, we evaluate a set of dense retrievers (e.g., Contriever, E5) at multiple stages of their training pipeline: in their purely unsupervised state, after fine-tuning on a general-purpose corpus like MS MARCO, and after fine-tuning on specific in-domain datasets (both human-written and LLM-generated). This multi-stage, multi-dataset evaluation framework is designed to answer the following key research questions:

\begin{itemize}
    \item[\textbullet] \textbf{RQ1:} Do unsupervised dense retrievers exhibit consistent preferences for LLM-generated content, or does this bias emerge primarily during supervised fine-tuning?
    \item[\textbullet] \textbf{RQ2:} How does supervised fine-tuning on various corpora (general corpora such as MS MARCO vs smaller in-domain datasets, in-domain human-written datasets vs. in-domain LLM-generated) affect the direction and magnitude of retrieval preferences?
    \item[\textbullet] \textbf{RQ3:} Is the prevailing perplexity-based explanation sufficient to account for the change in observed bias?
\end{itemize}

\subsection{Datasets and Metrics}

We use the SciFact \cite{DBLP:conf/emnlp/ScifactWaddenLKCBWH22} and NQ320K \cite{DBLP:journals/tacl/NQKwiatkowskiPRCP19} datasets following Dai et al. \cite{DBLP:conf/kdd/DaiNeuralZPLHLZW024}, employing their provided parallel corpora of human-written and Llama2-generated passages that maintain semantic equivalence.
To quantify retrieval preferences, we follow Dai et al.'s approach using the Relative $\Delta$ metric: Human-written and LLM-generated passages are merged into a single corpus for retrieval. NDCG scores are then calculated separately for each source type within the same ranked list, treating documents from the opposite source as irrelevant while preserving the original ranking order. These scores are synthesized as:

\begin{equation*}
\text{Relative } \Delta =
\frac{\text{NDCG}_{\text{Human}} - \text{NDCG}_{\text{LLM}}}
     {\text{NDCG}_{\text{Human}} + \text{NDCG}_{\text{LLM}}}
\times 100\%
\end{equation*}

Positive Relative $\Delta$ indicate a preference for human-written content, while negative values indicate a preference for LLM-generated content. Dataset statistics are summarized in Table~\ref{tab:dataset_stats}.

\begin{table}[t]
\centering
\caption{Dataset statistics for the original human-written and the Llama2-generated corpora. Avg. Rel Docs is the number of average relevant documents per query.}
\label{tab:dataset_stats}
\setlength{\tabcolsep}{4pt} 
\begin{tabular}{ll rr rr c}
\toprule
Dataset & Corpus & \shortstack{\# Train \\ Pairs} & \shortstack{\# Test \\ Queries} & \# Docs & \shortstack{Avg. Doc \\ Length} & \shortstack{Avg. Rel\\ Docs} \\
\midrule
\multirow{2}{*}{SciFact} & Human-Written & \multirow{2}{*}{1,408} & \multirow{2}{*}{300} & 5,183 & 201.81 & 1.1 \\
 & Llama2-Gen. & & & 5,183 & 192.66 & 1.1 \\
\midrule
\multirow{2}{*}{NQ320K} & Human-Written & \multirow{2}{*}{307,373} & \multirow{2}{*}{7,830} & 109,739 & 199.79 & 1.0 \\
 & Llama2-Gen. & & & 109,739 & 174.49 & 1.0 \\
\bottomrule
\end{tabular}
\end{table}

\subsection{Dense Retriever Selection}

Our study investigates the source bias of a variety of dense retriever families: 
E5, Contriever, and AugTriever. We selected these models because unlike older 
models (e.g., ANCE, TAS-B), they provide effective unsupervised checkpoints 
that function independently of MS MARCO. Models like ANCE \cite{DBLP:ANCEconf/iclr/XiongXLTLBAO21} and \mbox{TAS-B} \cite{DBLP:TASBconf/sigir/HofstatterLYLH21} are 
typically initialized from BERT and immediately fine-tuned on MS MARCO, making 
it impossible to separate the effect of the model architecture from the fine-tuning dataset. 
By using E5, Contriever, and AugTriever, we can evaluate the model's bias in its unsupervised 
state first, and then clearly observe how introducing MS MARCO alters that 
preference. Below we summarize the models used in our experiments.

\begin{itemize}
    \item[\textbullet] \textbf{E5} \cite{DBLP:journals/corr/E5abs-2212-03533} is a general-purpose embedding model trained in two stages: first, unsupervised contrastive pre-training on a large corpus of web-mined text pairs, followed by supervised fine-tuning on a mixture of datasets including NLI and MS MARCO.
    
    \item[\textbullet] \textbf{Contriever} \cite{DBLP:journals/tmlr/IzacardCHRBJG22} is a dense retriever also trained with a two-stage process, beginning with unsupervised contrastive learning on Wikipedia and CCNet, followed by supervised fine-tuning on MS MARCO.
    
    \item[\textbullet] \textbf{AugTriever} \cite{meng2022augtriever} is a family of purely unsupervised dense retrievers that learn from synthetically generated query-document pairs instead of human-labeled data. Queries are automatically created from passages using techniques like Transferred Query Generation (TQGEN) \cite{meng2022augtriever}, and the resulting pairs are used to pre-train a BERT-base bi-encoder via contrastive learning. We use two variants from this family: \textbf{Hybrid-TQGen+} and \textbf{Hybrid-All}.
\end{itemize}

\subsection{Fine-tuning and Implementation Details}\label{sec:finetuning_details}

To systematically analyze how retrieval preferences evolve, we conduct fine-tuning 
experiments on three distinct types of corpora: 1) the large, general-purpose 
MS MARCO dataset, 2) smaller in-domain datasets (NQ320K and SciFact), and 3) 
the LLM-generated versions of these in-domain datasets. All experiments were implemented using the Hugging Face Sentence Transformers library \cite{DBLP:conf/emnlp/ReimersG19}. For all fine-tuning runs, we employed a consistent training setup. The models were trained in a distributed setting using DeepSpeed across four GPUs with the standard InfoNCE contrastive loss \cite{DBLP:journals/corr/oordabs-1807-03748} as the objective. We kept the core hyperparameters constant: the learning rate was set to $1 \times 10^{-5}$, the per-device batch size was 16 with 4 gradient accumulation steps (for an effective batch size of 256), and a warmup ratio of 0.05 was applied.

Our specific experimental conditions and the starting checkpoints for each model are detailed below:
\begin{itemize}
    \item[\textbullet] \textbf{MS MARCO Fine-tuning:} To analyze the effect of fine-tuning on this common, large-scale dataset, we started from the purely unsupervised checkpoints of E5 and the AugTriever variants and fine-tuned them for 1 epoch. For Contriever, we used the standard, publicly available checkpoint already fine-tuned on MS MARCO from the Sentence Transformers library.
    
    \item[\textbullet] \textbf{In-Domain Fine-tuning (SciFact \& NQ320K):} For these experiments, we started from the unsupervised checkpoints for all models. For SciFact, we trained for 30 epochs to ensure convergence. For NQ320K, which contains 307,373 training pairs, we sampled 30,000 pairs for training and fine-tuned over 3 epochs.
    
    \item[\textbullet] \textbf{LLM-Corpus Fine-tuning:} We fine-tuned models on the 
    `LLM-generated' versions of the corpora. We utilize the parallel corpora provided 
    by Dai et al. \cite{DBLP:conf/kdd/DaiNeuralZPLHLZW024}, where passages from SciFact and NQ320K were rewritten by \mbox{Llama-2} to maintain semantic equivalence to the human gold passages. The same hyperparameters (epochs/sampling) as the standard in-domain experiments were used.
\end{itemize}

%% file: results_and_analysis.tex
\section{Results and Analysis}
\label{sec:results}

This section presents the results of our experimental evaluation.

\subsection{Revisiting Source Bias in Dense Retrievers (RQ1)}

We begin by reproducing the main results from prior work with the 4 dense retrievers: ANCE \cite{DBLP:ANCEconf/iclr/XiongXLTLBAO21}, BERM \cite{DBLP:BERMconf/acl/XuPSC23}, TAS-B \cite{DBLP:TASBconf/sigir/HofstatterLYLH21}, and Contriever \cite{DBLP:journals/tmlr/IzacardCHRBJG22} (the checkpoint that was also finetuned on MS-MARCO). Table~\ref{tab:original_bias_results_ndcg} reproduces the key \textit{source bias} result from Dai et al. \cite{DBLP:conf/kdd/DaiNeuralZPLHLZW024}. We see that across all the dense retrievers, the Relative~$\Delta$ (human vs. LLM) is consistently negative on both SciFact and NQ320K, indicating a pro-LLM preference.

\begin{table}[t]
\centering
\caption{Performance comparison (NDCG) for neural retrieval models reproducing the key "source bias" results from Dai et al.  \cite{DBLP:conf/kdd/DaiNeuralZPLHLZW024} for ANCE, BERM, TAS-B, and Contriever MS-MARCO. Negative values in the "Relative $\Delta$" row, highlighted in blue, indicate a bias towards LLM-generated content.\protect\footnotemark}
\label{tab:original_bias_results_ndcg}
\setlength{\tabcolsep}{4pt}
\begin{tabular}{ll *{6}{c}}
\toprule
\multirow{2}{*}{Model} & \multirow{2}{*}{Corpus} & \multicolumn{3}{c}{SciFact} & \multicolumn{3}{c}{NQ320K} \\
\cmidrule(lr){3-5} \cmidrule(lr){6-8}
& & \shortstack{NDCG \\ @1} & \shortstack{NDCG \\ @3} & \shortstack{NDCG \\ @5} & \shortstack{NDCG \\ @1} & \shortstack{NDCG \\ @3} & \shortstack{NDCG \\ @5} \\
\midrule
\multirow{3}{*}{ANCE} & Human & 15.3 & 30.1 & 32.7 & 22.2 & 41.2 & 44.6 \\
& LLM-Gen. & 24.7 & 35.8 & 37.7 & 29.1 & 45.9 & 49.0 \\
& Rel. $\Delta$ & \cellcolor{BiasBlue}-47.0 & \cellcolor{BiasBlue}-17.3 & \cellcolor{BiasBlue}-14.2 & \cellcolor{BiasBlue}-26.9 & \cellcolor{BiasBlue}-10.8 & \cellcolor{BiasBlue}-9.4 \\
\midrule
\multirow{3}{*}{BERM} & Human & 16.3 & 30.2 & 31.8 & 18.6 & 37.5 & 40.7 \\
& LLM-Gen. & 23.7 & 34.1 & 36.4 & 31.6 & 47.0 & 50.0 \\
& Rel. $\Delta$ & \cellcolor{BiasBlue}-37.0 & \cellcolor{BiasBlue}-12.1 & \cellcolor{BiasBlue}-13.5 & \cellcolor{BiasBlue}-51.8 & \cellcolor{BiasBlue}-22.5 & \cellcolor{BiasBlue}-20.5 \\
\midrule
\multirow{3}{*}{TAS-B} & Human & 20.0 & 40.2 & 43.1 & 25.7 & 45.4 & 48.8 \\
& LLM-Gen. & 31.7 & 44.8 & 47.5 & 27.6 & 46.5 & 50.0 \\
& Rel. $\Delta$ & \cellcolor{BiasBlue}-45.3 & \cellcolor{BiasBlue}-10.8 & \cellcolor{BiasBlue}-9.7 & \cellcolor{BiasBlue}-7.1 & \cellcolor{BiasBlue}-2.4 & \cellcolor{BiasBlue}-2.4 \\
\midrule
\multirow{3}{*}{\shortstack{Contriever \\ MS-MARCO}} & Human & 24.0 & 43.7 & 47.8 & 25.9 & 48.5 & 51.9 \\
& LLM-Gen. & 31.0 & 47.8 & 50.5 & 32.5 & 51.9 & 55.4 \\
& Rel. $\Delta$ & \cellcolor{BiasBlue}-25.5 & \cellcolor{BiasBlue}-9.0 & \cellcolor{BiasBlue}-5.5 & \cellcolor{BiasBlue}-22.6 & \cellcolor{BiasBlue}-6.8 & \cellcolor{BiasBlue}-6.5 \\
\bottomrule
\end{tabular}
\end{table}

\footnotetext{MAP results are omitted for brevity as they exhibit the same bias trend. BERM parameter details were not released so they were copied over from \cite{DBLP:conf/kdd/DaiNeuralZPLHLZW024}. }

\begin{table}[t]
\centering
\caption{Results for additional dense retrievers. Unsupervised checkpoints (E5-Base, Contriever, and AugTriever variants) show mixed source preferences rather than a consistent pro-LLM bias. The \textit{Relative~$\Delta$} row with cells shaded in blue indicates pro-LLM preferences while cells shaded in orange indicate pro-human preferences. The same color scheme is used in subsequent tables.}

\label{tab:extended_bias_results}
\setlength{\tabcolsep}{4pt} 
\begin{tabular}{ll *{6}{c}}
\toprule
\multirow{2}{*}{Model} & \multirow{2}{*}{Corpus} & \multicolumn{3}{c}{SciFact} & \multicolumn{3}{c}{NQ320K} \\
\cmidrule(lr){3-5} \cmidrule(lr){6-8}
& & \shortstack{NDCG \\ @1} & \shortstack{NDCG \\ @3} & \shortstack{NDCG \\ @5} & \shortstack{NDCG \\ @1} & \shortstack{NDCG \\ @3} & \shortstack{NDCG \\ @5} \\
\midrule
\multirow{3}{*}{\shortstack{E5-Base \\ Unsupervised}} & Human & 5.3 & 17.6 & 20.6 & 33.2 & 49.3 & 52.6 \\
& LLM-Gen. & 20.3 & 27.6 & 30.9 & 19.7 & 39.7 & 43.7 \\
& Rel. $\Delta$ & \cellcolor{BiasBlue}-117.2 & \cellcolor{BiasBlue}-44.2 & \cellcolor{BiasBlue}-40.0 & \cellcolor{BiasOrange}51.0 & \cellcolor{BiasOrange}21.6 & \cellcolor{BiasOrange}18.5 \\
\midrule
\multirow{3}{*}{E5} & Human & 26.0 & 48.5 & 50.9 & 26.1 & 49.6 & 53.2 \\
& LLM-Gen. & 32.7 & 50.3 & 53.7 & 35.1 & 54.8 & 58.2 \\
& Rel. $\Delta$ & \cellcolor{BiasBlue}-22.7 & \cellcolor{BiasBlue}-3.7 & \cellcolor{BiasBlue}-5.4 & \cellcolor{BiasBlue}-29.4 & \cellcolor{BiasBlue}-9.9 & \cellcolor{BiasBlue}-9.0 \\
\midrule
\multirow{3}{*}{Contriever} & Human & 27.7 & 43.3 & 47.1 & 23.5 & 38.8 & 42.6 \\
& LLM-Gen. & 22.7 & 41.7 & 44.1 & 19.6 & 36.1 & 40.1 \\
& Rel. $\Delta$ & \cellcolor{BiasOrange}19.9 & \cellcolor{BiasOrange}3.8 & \cellcolor{BiasOrange}6.5 & \cellcolor{BiasOrange}17.7 & \cellcolor{BiasOrange}7.1 & \cellcolor{BiasOrange}6.0 \\
\midrule
\multirow{3}{*}{\shortstack{AugTriever \\ Hybrid TQGen+}} & Human & 21.7 & 36.5 & 39.1 & 7.7 & 18.2 & 21.5 \\
& LLM-Gen. & 24.3 & 35.3 & 38.7 & 21.8 & 31.1 & 34.1 \\
& Rel. $\Delta$ & \cellcolor{BiasBlue}-11.3 & \cellcolor{BiasOrange}3.3 & \cellcolor{BiasOrange}1.0 & \cellcolor{BiasBlue}-95.6 & \cellcolor{BiasBlue}-52.3 & \cellcolor{BiasBlue}-45.3 \\
\midrule
\multirow{3}{*}{\shortstack{AugTriever\\Hybrid All}} & Human & 24.7 & 39.4 & 41.1 & 10.6 & 25.6 & 29.6 \\
& LLM-Gen. & 22.7 & 37.5 & 40.8 & 28.4 & 40.7 & 43.9 \\
& Rel. $\Delta$ & \cellcolor{BiasOrange}8.4 & \cellcolor{BiasOrange}4.9 & \cellcolor{BiasOrange}0.7 & \cellcolor{BiasBlue}-91.3 & \cellcolor{BiasBlue}-45.6 & \cellcolor{BiasBlue}-38.9 \\
\bottomrule
\end{tabular}
\end{table}

What is notable about the results and choice of dense retrievers is the fact that  all the models were finetuned on MS-MARCO which potentially makes it a confounder. Furthermore, Contriever MS-MARCO was trained in two stages, and we believe the unsupervised checkpoint should have been included to see what the results would have been for a purely unsupervised trained model. Therefore, we have decided to include E5, both its unsupervised checkpoint and final finetuned checkpoint, Contriever only after its unsupervised checkpoint, and the two AugTriever variants' unsupervised checkpoints

Table~\ref{tab:extended_bias_results} presents the results for these additional dense retrievers. In contrast to Table~\ref{tab:original_bias_results_ndcg}, these models do not exhibit a uniform pro-LLM preference: signs of Relative~$\Delta$ vary by dataset and model, and several unsupervised models show clear pro-human bias on at least one benchmark (e.g., \textbf{Contriever (Unsupervised)} with $+19.9$ on SciFact+AIGC and $+17.7$ on NQ320K+AIGC).

The results presented in Table~\ref{tab:extended_bias_results} challenge the previously held conclusion that dense retrievers are natively biased towards LLM-generated content. We observe that models having undergone only unsupervised training do not exhibit a consistent bias; instead, the preference is highly dataset-dependent, as shown by the varied signs of the Relative~$\Delta$ values. For instance, \textbf{Contriever} (unsupervised only trained version) shows a consistent and strong preference for human-written documents, with a Relative~$\Delta$ on NDCG@1 of $+19.9$ on SciFact and $+17.7$ on NQ320K. \textbf{E5-Base Unsupervised} displays a more extreme but dataset-dependent bias: on SciFact it exhibits a pronounced pro-LLM bias with Relative~$\Delta$ of $-117.2$ at NDCG@1, yet on NQ320K the bias reverses to strongly favor human-written text, reaching $+51.0$ at NDCG@1. After supervised fine-tuning phase, the full \textbf{E5} model exhibits bias towards LLM-generated text but with more tempered Relative $\Delta$ values of $-22.7$ (SciFact) and $-29.4$ (NQ320K) at NDCG@1. The \textbf{AugTriever} family also exhibits dataset-dependent biases. \textbf{Hybrid TQGen+} shows a small pro-LLM bias on SciFact at NDCG@3 and NDCG@5  ($+3.3$ and $+1.0$ ) but flips dramatically on NQ320K with a large pro-LLM preference ($-52.3$ and $-45.3$). \textbf{Hybrid All} also produce similar numbers at NDCG@3 and NDCG@5. These contrasting results underscore that unsupervised dense retrievers do not converge on a uniform source preference; rather, their biases are shaped by the unsupervised training method itself, including choices of data and unsupervised training methods.

Taken together, Tables~\ref{tab:original_bias_results_ndcg} and \ref{tab:extended_bias_results} suggest that the pro-LLM bias is not an inherent property of dense retrievers. Rather, it appears closely tied to training configuration: the consistently biased models are those subjected to large-scale supervised fine-tuning (e.g., MS~MARCO), whereas only unsupervised training shows mixed—and sometimes opposite—preferences. This observation motivates RQ2, where we isolate the role of supervised fine-tuning in shaping source bias. 

This demonstrates that fine-tuning causes models to overfit to the stylistic features of the training corpus, creating a strong but brittle in-domain preference that fails to generalize.

\subsection{Examining Fine-tuning Effects on Source Bias (RQ2)}

\subsubsection{MS Marco Finetuning}

\definecolor{BiasBlue}{HTML}{D6EAF8}
\definecolor{BiasOrange}{HTML}{FDEBD0}

\begin{table}[t]
\centering
\caption{The effect of supervised fine-tuning on MS MARCO. Unlike the mixed results of unsupervised models (Table~\ref{tab:extended_bias_results}), fine-tuning on MS MARCO consistently induces a strong bias towards LLM-generated content.}

\label{tab:finetuning_results}
\setlength{\tabcolsep}{4pt}
\begin{tabular}{ll *{6}{c}}
\toprule
\multirow{2}{*}{Model} & \multirow{2}{*}{Corpus} & \multicolumn{3}{c}{SciFact} & \multicolumn{3}{c}{NQ320K} \\
\cmidrule(lr){3-5} \cmidrule(lr){6-8}
& & \shortstack{NDCG \\ @1} & \shortstack{NDCG \\ @3} & \shortstack{NDCG \\ @5} & \shortstack{NDCG \\ @1} & \shortstack{NDCG \\ @3} & \shortstack{NDCG \\ @5} \\
\midrule
\multirow{3}{*}{\shortstack{E5-Base \\ Unsupervised}} & Human & 9.3 & 17.4 & 19.4 & 11.2 & 19.5 & 21.7 \\
& LLM-Gen. & 13.7 & 18.8 & 21.5 & 13.3 & 21.3 & 23.4 \\
& Rel. $\Delta$ & \cellcolor{BiasBlue}-37.7 & \cellcolor{BiasBlue}-7.5 & \cellcolor{BiasBlue}-10.2 & \cellcolor{BiasBlue}-17.2 & \cellcolor{BiasBlue}-8.5 & \cellcolor{BiasBlue}-7.9 \\
\midrule
\multirow{3}{*}{Contriever} & Human & 24.0 & 43.7 & 47.8 & 25.9 & 48.5 & 51.9 \\
& LLM-Gen. & 31.0 & 47.8 & 50.5 & 32.5 & 51.9 & 55.4 \\
& Rel. $\Delta$ & \cellcolor{BiasBlue}-25.5 & \cellcolor{BiasBlue}-9.0 & \cellcolor{BiasBlue}-5.5 & \cellcolor{BiasBlue}-22.6 & \cellcolor{BiasBlue}-6.8 & \cellcolor{BiasBlue}-6.5 \\
\midrule
\multirow{3}{*}{\shortstack{AugTriever \\ Hybrid TQGen+}} & Human & 3.0 & 6.1 & 6.7 & 9.6 & 15.4 & 17.1 \\
& LLM-Gen. & 6.7 & 8.7 & 9.8 & 8.9 & 15.3 & 17.1 \\
& Rel. $\Delta$ & \cellcolor{BiasBlue}-75.9 & \cellcolor{BiasBlue}-34.2 & \cellcolor{BiasBlue}-38.1 & \cellcolor{BiasOrange}7.4 & \cellcolor{BiasOrange}0.8 & \cellcolor{BiasOrange}0.2 \\
\midrule
\multirow{3}{*}{\shortstack{AugTriever \\ Hybrid-All}} & Human & 3.0 & 5.5 & 6.3 & 8.7 & 14.5 & 16.4 \\
& LLM-Gen. & 6.3 & 9.5 & 9.8 & 9.2 & 15.3 & 17.2 \\
& Rel. $\Delta$ & \cellcolor{BiasBlue}-71.4 & \cellcolor{BiasBlue}-53.0 & \cellcolor{BiasBlue}-44.5 & \cellcolor{BiasBlue}-5.3 & \cellcolor{BiasBlue}-5.2 & \cellcolor{BiasBlue}-4.5 \\
\bottomrule
\end{tabular}
\end{table}
As MS MARCO is the most widely used dataset for supervised fine-tuning dense retrievers, we begin by isolating its effect. Specifically, we fine-tune the unsupervised checkpoints of Contriever, E5, and the two AugTriever variants, and then analyze how their source preferences change.

The results in Table~\ref{tab:finetuning_results} confirm that supervised fine-tuning on MS MARCO can heavily alter the source preference of a model, often reversing its original preference entirely. The most notable example is \textbf{Contriever}. In its unsupervised state (Table~\ref{tab:extended_bias_results}), it exhibited a strong, consistent bias towards human-written content, with a Relative~$\Delta$ on NDCG@1 of $+19.9$ on SciFact and $+17.7$ on NQ320K. After fine-tuning on MS MARCO, this bias completely flips to a strong preference for LLM-generated content, with the Relative~$\Delta$ decreasing to $-25.5$ on SciFact and $-22.6$ on NQ320K.

A notable observation is that fine-tuning on MS MARCO can also temper the magnitude of a pre-existing bias while still keeping its direction intact. For example, \textbf{E5 Unsupervised} already had a strong pro-LLM bias on SciFact with a Relative~$\Delta$ of $-117.2$ on NDCG@1. After fine-tuning, this bias, while still strongly pro-LLM, was reduced to $-37.7$.

Ultimately, while the magnitude of the Relative~$\Delta$ varies, the most critical factor for ranking is the direction of the bias. In a retrieval context, if an LLM-generated document receives even a slightly higher relevance score than its human-written counterpart, it will be ranked higher. The key takeaway is that fine-tuning on MS MARCO makes a pro-LLM ranking outcome much more likely, regardless of the model's initial disposition. This raises an important question: does supervised fine-tuning in general drive retrievers toward LLM-generated content, or is this effect specific to MS MARCO? To explore this, we next fine-tune on smaller in-domain datasets, SciFact and NQ320K, to examine how their source preferences evolve.

\subsubsection{In-Domain Finetuning}

\begin{table}[t]
\centering
\caption{Effect of in-domain fine-tuning on source bias. For the SciFact columns, models are fine-tuned on SciFact; for the NQ320K columns, models are fine-tuned on NQ320K. Fine-tuning on NQ320K yields a consistent pro-human bias, while fine-tuning on SciFact produces mixed results across models.}

\label{tab:in_domain_tuning_results}
\setlength{\tabcolsep}{4pt}
\begin{tabular}{ll *{6}{c}}
\toprule
\multirow{2}{*}{Model} & \multirow{2}{*}{Corpus} & \multicolumn{3}{c}{SciFact} & \multicolumn{3}{c}{NQ320K} \\
\cmidrule(lr){3-5} \cmidrule(lr){6-8}
& & \shortstack{NDCG \\ @1} & \shortstack{NDCG \\ @3} & \shortstack{NDCG \\ @5} & \shortstack{NDCG \\ @1} & \shortstack{NDCG \\ @3} & \shortstack{NDCG \\ @5} \\
\midrule
\multirow{3}{*}{\shortstack{E5-Base \\  Unsupervised}} & Human & 25.3 & 43.4 & 47.2 & 24.2 & 36.4 & 39.2 \\
& LLM-Gen. & 28.0 & 43.1 & 47.2 & 15.5 & 30.8 & 34.1 \\
& Rel. $\Delta$ & \cellcolor{BiasBlue}-10.0 & \cellcolor{BiasOrange}0.6 & \cellcolor{BiasBlue}-0.1 & \cellcolor{BiasOrange}43.7 & \cellcolor{BiasOrange}16.5 & \cellcolor{BiasOrange}14.0 \\
\midrule
\multirow{3}{*}{Contriever} & Human & 26.0 & 46.2 & 49.6 & 41.7 & 56.7 & 59.8 \\
& LLM-Gen. & 30.7 & 48.0 & 52.5 & 14.6 & 38.0 & 42.7 \\
& Rel. $\Delta$ & \cellcolor{BiasBlue}-16.5 & \cellcolor{BiasBlue}-3.7 & \cellcolor{BiasBlue}-5.7 & \cellcolor{BiasOrange}96.4 & \cellcolor{BiasOrange}39.5 & \cellcolor{BiasOrange}33.4 \\
\midrule
\multirow{3}{*}{\shortstack{AugTriever \\ Hybrid TQGen+}} & Human & 25.3 & 39.9 & 42.4 & 20.1 & 30.1 & 32.8 \\
& LLM-Gen. & 21.3 & 38.1 & 40.3 & 12.5 & 24.4 & 27.4 \\
& Rel. $\Delta$ & \cellcolor{BiasOrange}17.1 & \cellcolor{BiasOrange}4.6 & \cellcolor{BiasOrange}5.2 & \cellcolor{BiasOrange}46.4 & \cellcolor{BiasOrange}21.1 & \cellcolor{BiasOrange}17.8 \\
\midrule
\multirow{3}{*}{\shortstack{AugTriever \\ Hybrid All}} & Human & 21.7 & 38.7 & 41.2 & 17.8 & 27.4 & 30.2 \\
& LLM-Gen. & 24.0 & 37.4 & 41.1 & 12.8 & 23.8 & 26.9 \\
& Rel. $\Delta$ & \cellcolor{BiasBlue}-10.2 & \cellcolor{BiasOrange}3.6 & \cellcolor{BiasOrange}0.2 & \cellcolor{BiasOrange}32.6 & \cellcolor{BiasOrange}14.0 & \cellcolor{BiasOrange}11.8 \\
\bottomrule
\end{tabular}
\end{table}

We now fine-tune the same four models that had only undergone unsupervised training (Contriever, E5, and the two AugTriever variants) on the in-domain datasets SciFact and NQ320K, following the setup described in Section~\ref{sec:finetuning_details}; Table~\ref{tab:in_domain_tuning_results} reports the results. On NQ320K, \textbf{all models shift heavily toward human-written passages}: Relative~$\Delta$ at NDCG@1 is +43.7 for E5-Base Unsupervised, +96.4 for Contriever, +46.4 for AugTriever Hybrid TQGen+, and +32.6 for AugTriever Hybrid All, with the same trend persisting at @3/@5. In contrast, when fine-tuned on SciFact, preferences are heterogeneous: Contriever and E5-Base Unsupervised show pro-LLM shifts ($-16.5$ and $-10.0$ at @1, respectively), AugTriever Hybrid TQGen+ is pro-human ($+17.1$), and AugTriever Hybrid All is slightly pro-LLM ($-10.2$); @3/@5 exhibit similar variability.

These results show that supervised in-domain fine-tuning does \emph{not} always cause the retriever to favor LLM-generated text; instead, the direction of source preference is both dataset- and model-dependent. This naturally raises the question of what happens when the in-domain training corpus itself is composed of LLM-generated passages. To investigate this, we fine-tune the same models from their unsupervised checkpoints on the LLM-generated counterparts of SciFact and NQ320K.

\subsubsection{In-Domain LLM-Generated Passage Finetuning}

\begin{table}[t]
\centering
\caption{Effect of fine-tuning on LLM-generated in-domain corpora. Across both datasets, fine-tuning consistently shifts models toward a pro-LLM bias, often reversing prior pro-human preferences (e.g., Contriever on NQ320K). This shows that exposure to synthetic text tends to systematically alter retrieval preferences.}

\label{tab:llm_corpus_tuning_results}

\setlength{\tabcolsep}{4pt}
\begin{tabular}{ll *{6}{c}}
\toprule
\multirow{2}{*}{Model} & \multirow{2}{*}{Corpus} & \multicolumn{3}{c}{SciFact} & \multicolumn{3}{c}{NQ320K} \\
\cmidrule(lr){3-5} \cmidrule(lr){6-8}
& & \shortstack{NDCG \\ @1} & \shortstack{NDCG \\ @3} & \shortstack{NDCG \\ @5} & \shortstack{NDCG \\ @1} & \shortstack{NDCG \\ @3} & \shortstack{NDCG \\ @5} \\
\midrule
\multirow{3}{*}{\shortstack{E5-Base \\ Unsupervised}} & Human & 22.0 & 39.8 & 42.6 & 19.2 & 32.7 & 35.9 \\
& LLM-Gen. & 26.0 & 40.2 & 44.4 & 20.3 & 33.8 & 37.0 \\
& Rel. $\Delta$ & \cellcolor{BiasBlue}-16.7 & \cellcolor{BiasBlue}-1.1 & \cellcolor{BiasBlue}-4.1 & \cellcolor{BiasBlue}-5.7 & \cellcolor{BiasBlue}-3.3 & \cellcolor{BiasBlue}-3.0 \\
\midrule
\multirow{3}{*}{{Contriever}} & Human & 20.0 & 42.1 & 46.2 & 22.1 & 46.1 & 50.0 \\
& LLM-Gen. & 36.0 & 51.0 & 55.2 & 36.7 & 54.5 & 58.1 \\
& Rel. $\Delta$ & \cellcolor{BiasBlue}-57.1 & \cellcolor{BiasBlue}-19.1 & \cellcolor{BiasBlue}-17.8 & \cellcolor{BiasBlue}-49.7 & \cellcolor{BiasBlue}-16.7 & \cellcolor{BiasBlue}-15.0 \\
\midrule
\multirow{3}{*}{\shortstack{AugTriever \\ Hybrid TQGen+}} & Human & 20.3 & 37.2 & 40.1 & 31.2 & 50.6 & 54.0 \\
& LLM-Gen. & 26.0 & 39.7 & 41.9 & 26.6 & 48.2 & 52.1 \\
& Rel. $\Delta$ & \cellcolor{BiasBlue}-24.5 & \cellcolor{BiasBlue}-6.6 & \cellcolor{BiasBlue}-4.3 & \cellcolor{BiasOrange}16.1 & \cellcolor{BiasOrange}5.0 & \cellcolor{BiasOrange}3.7 \\
\midrule
\multirow{3}{*}{\shortstack{AugTriever \\ Hybrid All}} & Human & 16.3 & 35.8 & 38.9 & 15.7 & 26.7 & 29.6 \\
& LLM-Gen. & 28.7 & 39.2 & 42.7 & 16.3 & 27.2 & 30.1 \\
& Rel. $\Delta$ & \cellcolor{BiasBlue}-54.8 & \cellcolor{BiasBlue}-9.1 & \cellcolor{BiasBlue}-9.4 & \cellcolor{BiasBlue}-3.6 & \cellcolor{BiasBlue}-1.9 & \cellcolor{BiasBlue}-1.4 \\
\bottomrule
\end{tabular}
\end{table}

The results in Table~\ref{tab:llm_corpus_tuning_results} show the effect of fine-tuning unsupervised checkpoints on LLM-generated corpora. Across both datasets, most models develop a strong preference for LLM-generated content. For example, Contriever, which exhibited a large pro-human bias when fine-tuned on the human-written NQ320K corpus (Relative~$\Delta$ of $+96.4$ in Table~\ref{tab:in_domain_tuning_results}), shifts to a strong pro-LLM preference when instead fine-tuned on the LLM-generated version of NQ320K (Relative~$\Delta$ of $-49.7$).  

This contrast between human- and LLM-trained variants reveals a clear and concerning pattern: while fine-tuning on human-written data can instill strong pro-human preferences, fine-tuning on LLM-generated data tends to systematically push models toward pro-LLM bias. The dramatic reversal observed in Contriever underscores the risks of using synthetic corpora for training. As LLM-generated content becomes more prevalent on the web, indiscriminate use of such text for retriever training may create a feedback loop where models increasingly learn to prefer AI-generated passages, potentially marginalizing human-written content in IR systems.

\subsection{RQ3: A Re-evaluation of the Perplexity Hypothesis}
\label{sec:rq3}

Our findings so far show that source bias is not a static property of dense retrievers but a dynamic one shaped by supervised fine-tuning. Prior work \cite{DBLP:conf/iclr/WangPerplexityDZP0WD0W25} attributed this bias to lower passage perplexity, measured as BERT masked language modeling (MLM) perplexity aggregated at the corpus level. However, such averaging conflates pairwise variation and does not capture how fine-tuning reshapes a specific retriever’s preferences.
We therefore evaluate perplexity from the model’s own perspective: if bias stems from a preference for low-perplexity text, then a fine-tuned retriever should assign higher relevance to passages it itself finds more predictable. 

\subsubsection{Measuring the Effects of Retriever-Centric Perplexity}

\begin{figure}[t!]
    \centering
    \includegraphics[width=\columnwidth]{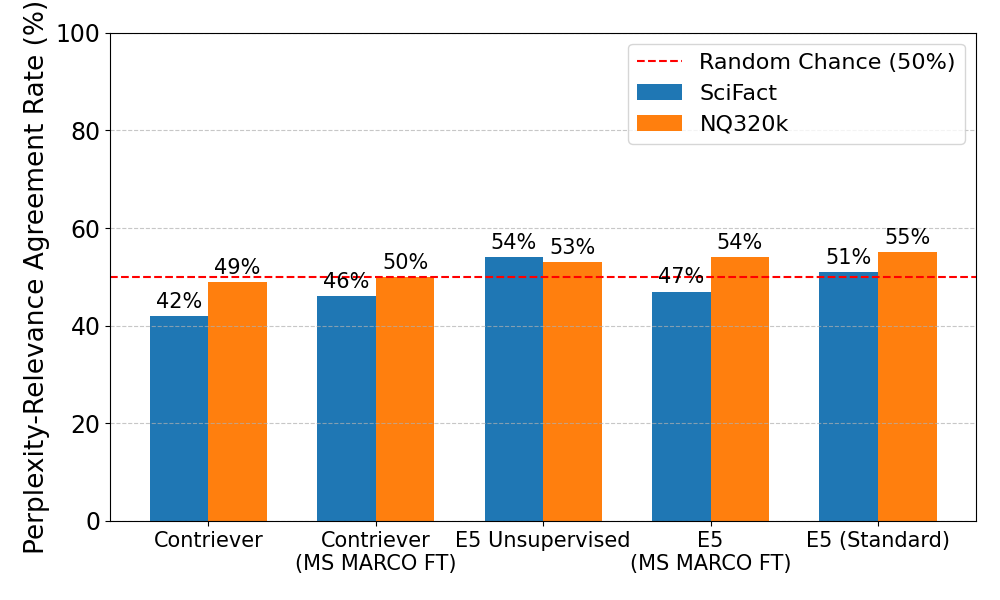}
    \caption{Perplexity–Relevance Agreement across training stages for E5 and Contriever. The dashed line marks 50\% (chance). While prior work hypothesized that lower perplexity passages should align with higher relevance scores, the agreement rates remain close to or below chance for all models and datasets, showing that perplexity even when measured from the retriever’s own encoder fails to account for the observed bias.}
    \label{fig:perplexity_agreement}
\end{figure}

\sloppy

To obtain a retriever-centric perplexity, we reattach a pre-trained BERT LM head to the fine-tuned retriever’s encoder \cite{DBLP:conf/iclr/WangPerplexityDZP0WD0W25}. Prior work \cite{DBLP:conf/iclr/WangPerplexityDZP0WD0W25} implicitly assumes bias is inherited from pretraining at the BERT training stage, leaving the effects of fine-tuning underexplored. Under that assumption, whether retriever-centric perplexity plays a role in a fine-tuned model’s relevance scoring was left unexamined. To test whether a retriever that prefers LLM-generated passages also finds them more predictable under its own representations, we introduce the \textbf{Perplexity-Relevance Agreement (PRA)} rate. To calculate this rate, we consider each query in our test set alongside its corresponding pair of a "gold" human-written passage and its semantically equivalent LLM-generated version. For each pair, we perform two calculations: we use the retriever to compute a relevance score for each passage with respect to the query, and we use our dense retriever with the language modeling head to calculate the retriever-centric perplexity of each passage. An \emph{agreement} is then counted if the passage with the lower retriever-centric perplexity also receives the higher relevance score. The PRA rate is the total fraction of agreements over all pairs, where 50\% indicates chance-level alignment, and a rate substantially greater than 50\% would support the perplexity hypothesis.

As shown in Figure~\ref{fig:perplexity_agreement}, agreement rates hover near the 50\% baseline across models and fine-tuning states and often fall below it. For example, Contriever on SciFact achieves only 42.2\%, indicating it more often favors the higher-perplexity passage; the MS~MARCO–tuned Contriever reaches just 46.1\% on the same dataset. These results indicate that even when measured from the retriever’s own perspective, perplexity does not explain the observed source bias. While human and LLM texts may differ in average predictability, dense retrievers do not consistently use such differences as a shortcut to determine passage relevance to a query.

%% file: conclusion.tex
\section{Conclusion}
\label{sec:conclusion}

In this paper, we reexamined the phenomenon of source bias in dense retrievers and demonstrated that it is not simply a preference for content generated by LLMs. Rather, the presence and direction of bias are shaped by training; different stages and datasets exert distinct effects. Our experiments demonstrated that fine-tuning on the popular MS MARCO corpus consistently induces a pro-LLM bias and that fine-tuning on LLM-generated corpora reliably reinforces this bias. In contrast, the effect of fine-tuning on human-written, in-domain corpora is less consistent. For instance, NQ320K produces a clear shift toward a pro-human preference, whereas SciFact shows no consistent bias. Additionally, we found that perplexity does not adequately explain these behaviors because even retriever-centric perplexity failed to predict relevance preferences. One remaining question is why some human-written corpora reliably alter dense retrievers toward LLM-generated content, while others do not. Future work should focus on explainability and interpretability techniques to identify the specific linguistic or distributional cues that dense retrievers internalize during fine-tuning.